\documentclass[aps,prb,floatfix,superscriptaddress,twocolumn]{revtex4}
\usepackage{graphics,latexsym}
\begin{document}

%\title{Excitation delocalization and the efficiency of energy transfer in a light-harvesting system}
\title{Efficiency of energy transfer in a light-harvesting system under quantum coherence}
\author{Alexandra Olaya-Castro}\email{a.olaya@physics.ox.ac.uk}
\author{Chiu Fan \surname{Lee}}
\author{Francesca Fassioli Olsen}
\affiliation{Department of Physics, University of Oxford, Clarendon Laboratory, Parks Road, OX1 3PU, United Kingdom}
\author{Neil F. Johnson}
\affiliation{Department of Physics, University of Miami, Memorial Drive, Coral Gables, Miami, Florida FL33126, USA}

%\date{\today}

\begin{abstract}
We investigate the role of quantum coherence in the efficiency of excitation transfer in a ring-hub arrangement
of interacting two-level systems, mimicking a light-harvesting antenna connected to a reaction center as it is found in natural photosynthetic systems. By using a quantum jump approach, we demonstrate that in the presence of quantum
coherent energy transfer and energetic disorder, the efficiency of excitation transfer from the antenna to the
reaction center depends intimately on the quantum superposition properties of the initial state. In particular, we
find that efficiency is sensitive to symmetric and asymmetric superposition of states in the basis of localized
excitations, indicating that initial state properties can be used as a efficiency control parameter at low
temperatures.

\end{abstract}

\maketitle
%%%%%%%%%%%%%%%%%%%%%%%%%%%%%%%%%%%%%%%%%%%%%%%%%%%%%%%%%%%%%%%%%%%%%%%%%
\section{Introduction}
%%%%%%%%%%%%%%%%%%%%%%%%%%%%%%%%%%%%%%%%%%%%%%%%%%%%%%%%%%%%%%%%%%%%%%%%%
Solar energy conversion in photosynthetic bacteria relies on
sophisticated light-harvesting (LH) antennae which capture photons and then
transfer the electronic excitation to a molecular complex which serves as a reaction centre (RC).
There charge separation takes place and chemical energy storage is initiated (e.g., see Ref. 1).
Some of the harvesting complexes (LH1) and the RC are closely associated and form a
core unit to ensure an efficient pathway for the transfer of excitations coming from peripheral
antennae (LH2). This transfer takes only a few hundred picoseconds and is
performed with extraordinarily high efficiency: most of the absorbed photons
give rise to a charge separation event \cite{hu02}. The precise mechanisms underlying such high efficiency remain
elusive despite numerous studies on the subject. In particular, whether quantum coherence plays any role on promoting
the efficiency is still ambiguous. Some works indicate that it
will induce higher excitation-transfer rates \cite{jang04}
while others argue that this may not necessarily be the case
\cite{gaab04}. Remarkably, recent experimental and theoretical works \cite{engel07, lee07, grondelle06} indicate that
long-lasting electronic coherence can indeed influence the excitation transfer dynamics in photosynthetic complexes.
For instance, quantum beats associated with electronic coherence in the Fenna-Matthews-Olson (FMO) complex of green
sulfur bacteria, which connects a large LH to the RC has been reported by Engel {\it et al.} \cite{engel07}. Also,
coherence among electronic states of closely associated pigments in the RC of purple bacteria has been reported by Lee
{\it et al.}\cite{lee07}. Furthermore, the excitation
transfer in organic dendrimers has  attracted
significant attention through the prospect of creating artificial photosynthetic systems \cite{andrews04}.
One of the key observations in these artificial systems is the evidence
of coherent energy transfer mechanisms. These experiments therefore open up the possibility of
exploring in detail the interplay between quantum coherence and the efficiency of natural and artificial LH
systems.
%%%%%%%%%%%%%%%%%%%%%%%  Fig 1 %%%%%%%%%%%%%%%%%%%%%%%%%%%%%%%%%%
\begin{figure}
\centering
\resizebox{6cm}{!}{\includegraphics*{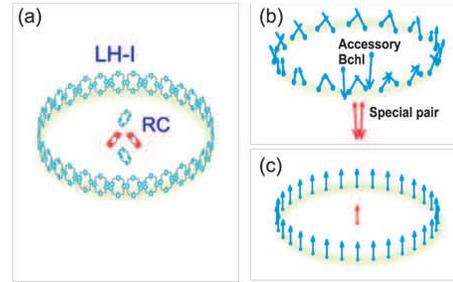}}
\caption{Schematic of the LH1-RC core of purple bacteria
Rodobacter Sphaeroides. (a) Arrangement of the 32 Bacterioclorophils (BChl)
surrounding  the RC. The RC has two accessory BChl  and two
acceptors forming a special pair responsible for charge
separation. (b) Diagram of the induced dipole
moments in (a). The arrows indicate the dipole moment directions
corresponding to data taken from Xhu {\it et al.} \cite{hu97}. (c) Toy
model: the RC is assumed to be a single two-level system.}
\label{fig:psu}
\end{figure}
%%%%%%%%%%%%%%%%%%%%%%% End fig 1 %%%%%%%%%%%%%%%%%%%%%%%%%%%%%%%%%

In this work we consider as model system a ring-hub arrangement of interacting two-level systems,
representing a LH1-RC core unit as in purple bacteria \cite{damjanovi00,hu97,hu98}, and use the quantum jump
approach \cite{carmichael93,plenio98} to provide a simple picture of how the quantum superposition properties of the
initial state of the excitation relate to the efficiency of transfer from the LH1 complex to the RC when coherent
energy transfer and static disorder of single-site energies dominate. The quantum jump approach
\cite{carmichael93,plenio98} proves to be particularly suitable to describe excitation dynamics in this situation
because the density matrix elements can be calculated exactly from the {\em no-jump evolution}, given that there is
only one excitation at most in the system and that only local dissipation rates (or charge separation rates) are
assumed \cite{hu02}. Considering initial states as superpositions of single-site excitations, we find that the efficiency profile depends both on the symmetry properties of such superposition states, and on the number
of sites among which the excitation is initially delocalized. In particular, our results show a non-trivial
interplay between excitation delocalization and efficiency as there can be an optimal delocalization length
for which efficiency of transfer reaches a maximum and transfer time a minimum. Such behaviour is robust
to the presence of energetic disorder. The plan of this paper is as follows. The next section outlines the model of energy transfer at very low
temperatures as well as how we use the quantum jump approach to calculate the main characteristics of our
photosynthetic system. Section III discusses our results both for a toy model and for a more detailed Hamiltonian
describing the LH1-RC complex.

%%%%%%%%%%%%%%%%%%%%%%%%%%%%%%%%%%%%%%%%%%%%%%%%%%%%%%%%%%%%%%%%%%%%%%%%%
\section{Coherent excitation transfer}
%%%%%%%%%%%%%%%%%%%%%%%%%%%%%%%%%%%%%%%%%%%%%%%%%%%%%%%%%%%%%%%%%%%%%%%%%
We consider a system of $M$ donor pigments surrounding a RC with
$N$ acceptors (see Fig. \ref{fig:psu}) described by the
Hamiltonian $H=H_0+H_I$. Labeling the donors from $1$ to $M$, and the acceptors from $M+1$ to $M+N$, the single
particle Hamiltonian is $H_0=\sum_{\alpha=1}^{M+N} \epsilon_{\alpha}\sigma^{+}_{\alpha}\sigma_{\alpha}^{-}$ where $\epsilon_{\alpha}$ is the
excitation energy of pigment $\alpha$, and the interaction Hamiltonian reads:
{\small
\begin{equation}
H_I=\sum_{j=1}^{M} \sum_{c=M+1}^{M+N}\gamma_{jc} \hat V_{jc}
+ \sum_{j=1; k>j}^{M} J_{jk} \hat V_{jk} + \sum_{c=M+1;r>c}^{M+N}g_{cr} \hat V_{cr} \ .
\label{eq:hi}
\end{equation}}Here $\hat V_{ab}=\sigma^{+}_{a}\sigma_{b}^{-}
+\sigma^{+}_{b}\sigma_{a}^{-}$ where $\sigma^{+(-)}$ is the Pauli operator for a two-level system, and
$\gamma_{jc}$, $J_{jk}$ and $g_{cr}$ are the donor-acceptor,
donor-donor, and acceptor-acceptor couplings, respectively.
We are interested in the low temperature regime where static disorder dominates and dynamical effects can be
neglected \cite{hofmann05}. In order to account for static disorder we treat $\epsilon_{\alpha}$ as having a
random component: $\epsilon_{\alpha}=E_{\alpha}(0) + \delta E_{\alpha}$ where $E_{\alpha}(0)$ is the ensemble
average value, and $\delta E_{\alpha}$ is the energy disorder at site $\alpha$ given by a Gaussian
distribution with zero mean and standard deviation $\sigma$.
We assume that the open system dynamics is dominated by
two incoherent processes: the excitation can be dissipated in a donor or it can induce charge separation at a site in the RC. Such  dynamics of our coherent photosynthetic core can be described by the Lindblad master
equation $(\hbar=1)$:
{\small
\begin{equation}
\frac{d}{dt}\rho = -i[H,\rho] +
\frac{1}{2}\sum_{\alpha}[2A_{\alpha}\rho A_{\alpha}^{\dag}-A_{\alpha}^{\dag}A_{\alpha}\rho -
\rho A_{\alpha}^{\dag}A_{\alpha}] \ ,
\label{eq:master}
\end{equation}}where the commutator generates the coherent part of the evolution and the action of
each operator $A_{\alpha}=\sqrt{2\Gamma_{\alpha}}\sigma_{\alpha}^{-}$ accounts for a
``jump" process associated either to dissipation of excitation in a donor
i.e $\Gamma_{\alpha}=\Gamma$ with $\alpha=1,\dots M$, or for a charge separation
event at an acceptor of the RC i.e $\Gamma_{\alpha}=\kappa$ with $\alpha=M+1,\dots M+N$. Here we have assumed
identical dissipation rates for the donors and identical charge separation rates for the acceptors at the RC. This
formalism can be extended to include other incoherent processes.

In order to solve equation (\ref{eq:master}) we follow the quantum jump approach \cite{carmichael93} and
re-write equation (\ref{eq:master}) as
\begin{equation}\label{eq:pcond}
\dot{\rho}  = -i(H_{cond}\rho-\rho H_{cond}^{\dag}) + \sum_{\alpha}A_{\alpha} \rho A_{\alpha}^\dag
\end{equation}
with
\begin{equation}
H_{cond}=H-i\Gamma\sum_{j=1}^{M}\sigma^{+}_{j}\sigma^{-}_{j}-
i\kappa\sum_{c=M+1}^{N}\sigma^{+}_{c}\sigma^{-}_{c} \ ,
\end{equation}
In this description, the excitation dynamics can be interpreted in terms of quantum trajectories where the system
follows a {\em no-jump evolution} associated to the non-hermitian Hamiltonian $H_{cond}$, interrupted by a single
stochastic collapse of the system to its ground state in the event of either dissipation or charge separation, with
probability $p_{\alpha}={\rm tr}[A_{\alpha}\rho A_{\alpha}^\dag]$. The {\it no-jump trajectory} conditioned on
no-decay-occur is described by $\dot{\rho}_{cond}(t)=-i(H_{cond} \rho-\rho H_{cond}^{\dag})$. In particular, if the
initial state is pure, i.e. $|\Psi(0)\rangle$, the state remains pure (but unnormalized, i.e. dissipative)
in the no-jump trajectory and becomes $|\Psi_{cond}(t)\rangle=exp(-i H_{cond}\, t) |\Psi(0)\rangle$.

We now demonstrate that given that a single excitation is present in our photosynthetic core, the dynamics of all the
density matrix elements can be calculated exactly knowing only $\dot{\rho}_{cond}(t)$. Notice that $H_{cond}$
preserves the number of excitations i.e. $[H_{cond},{\mathcal N}]=0$ with ${\mathcal
N}=\sum_{\alpha=1}^{M+N}\sigma^{+}_{\alpha}\sigma^{-}_{\alpha}$. Hence, for a single excitation the density matrix
dynamics is restricted to the subspace of single-excitation states plus the ground state. We choose the basis given
by  $S=\{|0\rangle, \{|j\rangle \},\{ |c\rangle \} \}$, where $|0\rangle=|0_1\dots 0_M;
0_{M+1}\dots 0_{M+N}\rangle$ is the state with all the pigments in their ground state, and
\begin{eqnarray*}
|j\rangle & = & |0_1\dots 1_j\dots 0_M;
0_{M+1}\dots 0_{M+N}\rangle  \\
|c\rangle & = &|0_1\dots0_M;0_{M+1}\dots 1_{c}\dots 0_{M+N}\rangle
\end{eqnarray*}
are states in which only the $j$th donor (or $c$th acceptor) is
excited. The labels after the semicolon in each ket refer to the acceptors at the RC. Let us denote the density matrix
elements $\rho_{kl}(t)=\langle k|\rho(t)|l\rangle$ for any pair of states $|k\rangle, |l\rangle \in S$. Notice that the second term of Eq. (\ref{eq:pcond}) satisfies
$\langle k |A_{\alpha} \rho(t) A_{\alpha}^\dag|l\rangle=0$ for all states except when $|k\rangle=|l\rangle=|0\rangle$. Therefore, for single-excitation states $\dot{\rho}_{kl}(t)=\langle k|\dot{\rho}_{cond}(t)|l\rangle$, while
$\dot{\rho}_{00}(t)=\sum_{\alpha}\langle 0|A_{\alpha} \rho(t)
A_{\alpha}^\dag|0\rangle=2\sum_{\alpha}\Gamma_{\alpha}\langle \alpha|\rho(t)|\alpha\rangle$. Now, since
$\rho(0)=\rho_{cond}(0)$, then $\dot{\rho}_{00}(t)=2\sum_{\alpha}\Gamma_{\alpha}\langle
\alpha|\rho_{cond}(t)|\alpha\rangle$. This demonstrates that the dynamics of all density matrix elements can be
entirely calculated with $\rho_{cond}(t)$ and hence our claim.
%%%%%%%%%%%%%%%%%%%%%%%%%%%%%%%%%%%%%%%%%%%%%%%%%%%%%%%%%%%%%%%%%%%%%%
\subsection{Efficiency and Transfer times}
%%%%%%%%%%%%%%%%%%%%%%%%%%%%%%%%%%%%%%%%%%%%%%%%%%%%%%%%%%%%%%%%%%%%%%%%
With the above formalism we can now focus on the main features of our coherent LH1-RC core. As it is
described in the review by Sener {\it et al.} \cite{sener05}, of particular interest are the efficiency, which is given by the probability of an excitation to be used for charge separation as opposed to being dissipated, the average
transfer time of excitation to get trapped by the RC, and the excitation lifetime after the initial absorption of a photon.

Let us denote the initial state $\Psi_0$. Clearly, the probability that the excitation is still in the system at time
$t$, i.e {\em no-jump probability} is $P(t;\Psi_0)=\sum_{k}^{M+N}\langle k|{\rho}_{cond}(t)|k\rangle$ while
$w(t;\Psi_0)=\dot{\rho}_{00}(t)$ is the probability density
that a `jump' (charge separation or dissipation) occurs between $[t,t+dt)$ and it reads
{\small
\begin{eqnarray}
w(t;\Psi_0)&=&2\Gamma\sum_{j=1}^M \langle j |\rho_{cond}(t)| j \rangle
+ 2\kappa\sum_{c=M+1}^{M+N} \langle c|\rho_{cond}(t)|c\rangle \nonumber \\
&\equiv& w_D(t;\Psi_0)+w_{RC}(t;\Psi_0)\ .
\end{eqnarray}}Here $w_D(t;\Psi_0)dt$ is the probability that it is dissipated
by any of the donors in  $[t,t+dt)$ while $w_{RC}(t;\Psi_0)dt$ is the
probability that the excitation is used for charge separation at the RC. 
Notice also that $w(t;\Psi_0)=-dP(t;\Psi_0)/dt$ leading to
$\int_0^{\infty}w(t;\Psi_0)dt=1$
which implies that the excitation will eventually either be
dissipated or trapped in the RC. In particular, for pure initial states of
the form $\Psi_0=\sum_{j=1}^M b_j(0)|j\rangle+ \sum_{c=M+1}^{M+N}b_c(0)|c\rangle$
with $\sum_{j=1}^M|b_j(0)|^2 + \sum_{c=M+1}^{M+N}|b_c(0)|^2=1$, we have that
$\rho_{cond}(t)=|{\Psi}_{cond}(t)\rangle\langle{\Psi}_{cond}(t)|$ with the unnormalized conditional state
given by
\begin{equation}\label{eq:state}
|{\Psi}_{cond}(t)\rangle=\sum_{j=1}^M b_{j}(t)|j\rangle~ + ~
\sum_{c=M+1}^{M+N} b_{c}(t)|c\rangle \ .
\end{equation}
The monotonically decreasing norm of this state gives the no-jump probability
$P(t;\Psi_0)=\parallel |{\Psi}_{cond}\rangle \parallel^2$, while
$w_D(t;\Psi_0)=2\Gamma \sum_{j=1}^M|{b}_j(t)|^2$, and
\begin{equation}\label{eq:wc}
w_{RC}(t;\Psi_0)=2\kappa\sum_{c=M+1}^{M+N}|{b}_{c}(t)|^2 \ .
\end{equation}
We therefore define the {\em efficiency} $(\eta)$ of energy transfer to the RC as the total probability that the
excitation is used in charge separation. The {\em transfer time}
$(t_f)$ is the average waiting-time before a jump associated with
charge-separation in the RC, given that the excitation was
initially in the LH1 ring. The {\em excitation lifetime} $(\tau)$ is the
average waiting-time before a jump of any kind occurs:
{\small
\[
\eta  = \int_0^{\infty}dt\,w_{RC}(t;\Psi_0) \ \ , \ \
t_f = \frac{1}{\eta}\int_0^{\infty}dt\,t\,w_{RC}(t;\Psi_0) \ ,
\]}
{\small
\begin{equation}
\tau = \int_0^{\infty}dt\,t\,w(t;\Psi_0) \ .
\end{equation}}
%%%%%%%%%%%%%%%%%%%%%%%%%%%%%%%%%%%%%%%%%%%%%%%%%%%%%%%%%%%%%%%%%%%%%%%%
\section{Efficiency control mechanisms}
%%%%%%%%%%%%%%%%%%%%%%%%%%%%%%%%%%%%%%%%%%%%%%%%%%%%%%%%%%%%%%%%%%%%%%%%
In the classical description \cite{sener05, ritz01}, where incoherent excitation transfer is assumed, typical efficiencies of
energy transfer tend to be near unit. This is due to a separation of the dissipation (ns) and the excitation transfer
and charge separation time scales (ps). We shall shortly show that using the same parameters for single-site energies,
electronic couplings, dissipation and charge separation rates, the efficiencies obtained under coherent transfer are
much lower and strongly dependent on the initial state of the excitation. We find that the initial relative phases between localized excitation states $|j\rangle$, and the number of donors among which the excitation is initially
delocalized, can act act as efficiency control mechanisms. In what follows, we first consider a simple model for which analytical solutions can be obtained and then we calculate efficiency and transfer times with the model Hamiltonian given by Hu {\it et al}. \cite{hu97,hu98}.

\subsection {A toy model for the LH1-RC complex}
The simplest model for which analytical solutions can be obtained corresponds to the RC taken as a single two-level
system i.e. $N=1$, on resonance with the $M=32$ donors in the LH1 ring (see Fig. \ref{fig:psu}(c)). Later we shall
show that the main qualitative behaviour observed in this situation also applies to a model featuring the detailed
structure of the LH1-RC complex in purple bacteria.
%%%%%%%%%%%%%%%%%%%%%%  Fig 2 %%%%%%%%%%%%%%%%%%%%%%%%%%%%%%%%%%
\begin{figure}
\centering
\resizebox{8cm}{!}{\includegraphics*{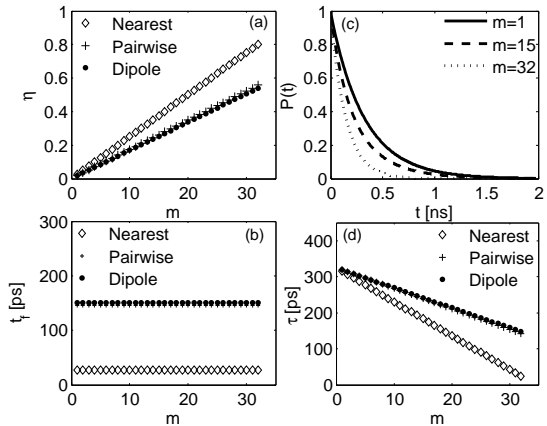}}
\caption{Numerical results for (a) $\eta$, (b) $t_f$, and
and (d) $\tau$ versus $m$,
for the toy model with three different interaction mechanisms. For
nearest-neighbour interactions ($\Diamond$) the coupling is 100meV,
while for the pairwise case ($+$) it is 10meV and equals the average
dipole-dipole coupling ($\bullet$). In each case, the donor-RC
coupling $\gamma$ equals 1meV,  $\Gamma=1\ \rm{ns}^{-1}$ and $\kappa=4\ \rm{ps}^{-1}$.
(c) No-jump probability  for the case of dipole-dipole interactions,
as a function of time and for different $m$ values.}
\label{fig:toy}
\end{figure}
%%%%%%%%%%%%%%%%%%%%%%% End fig 2 %%%%%%%%%%%%%%%%%%%%%%%%%%%%%%%%%
We consider initial states in which the excitation is delocalized among donors, i.e. $\Psi_0=\sum_{j=1}^M
b_j(0)|j\rangle$ with $\sum_{j=1}^M|b_j(0)|^2=1$. The unnormalized state of Eq. (\ref{eq:state}) becomes
$|\Psi_{cond}(t)\rangle=\sum_{j=1}^M {b_j}(t)|j\rangle~ + ~
b_{M+1}(t)|M+1\rangle$ satisfying the equation $d|{\Psi}_{cond}(t)\rangle/dt=-iH_{cond}|\Psi_{cond}(t)\rangle$ which leads
to a set of first-order coupled differential
equations for the complex amplitudes $b_j(t)$ and $b_{M+1}(t)$. In Appendix A we show that when the system's dynamics
is invariant with respect to exchange of donors in the antenna, one can find analytical solutions for $b_{M+1}(t)$.
Let us define the effective interaction between a donor $j$ and the rest of pigments in the LH1 ring as
$\Delta_j=\sum_{k}J_{jk}$ i.e. the sum of all the coupling strengths between donor $j$ and any other pigment in the
LH ring. The system's dynamics is invariant with respect to donor exchange when both all donor-RC couplings
are identical, i.e. $\gamma_{jc}\equiv\gamma$, and the effective interaction between a donor and the rest of pigments in the
ring are also identical for all donors, i.e. $\Delta_j=\Delta$ for all $j$. Under these conditions $b_{M+1}(t)$
satisfies the differential equation
$\ddot{b}_{M+1}(t) + X \dot{b}_{M+1}(t) + Y {b}_{M+1}(t) = 0$ with $X=(\kappa+\Gamma +i\Delta)$ and $Y=4M\gamma^2 +
\kappa (\Gamma+i\Delta)$ (see details in Appendix A). For the initial condition where $b_{M+1}(0)=0$ i.e. excitation is initially in the ring, we
find
\begin{equation}
|b_{M+1}(t)|^2={\mathcal F}(t)\big |\sum_{j=1}^M b_j(0)\big |^2  \ .
\end{equation}
Here ${\mathcal F}(t)=4\gamma^2 e^{-(\Gamma+\kappa)t}|{\rm sin}(\Omega
t/2)|^2/|\Omega|^2$ and $\Omega$ is the complex
frequency that determines the timescale of coherent oscillations i.e.
$\Omega=\sqrt{4M\gamma^2-(\Gamma-\kappa+i\Delta)^2}$. Note
that $\Delta$ identical for all donors does not imply that the pair couplings $J_{jk}$ need to
be identical for all possible pairs. Hence, analytical solutions can be found for three different mechanisms of
interaction between the donors: (i) nearest neighbours with
$J_{jj+1}=J/2$ (ii) pairwise interaction with
$J_{jk}\equiv J$ for all $\langle j,k \rangle$ pairs and (iii) dipole-dipole
interactions of the form $J_{jk}=J/r_{jk}^3$ where ${\boldmath \rm
r}_{jk}$ is the relative position vector between the induced
dipole moments of donors $j$ and $k$. From equation (\ref{eq:wc}) the probability density of having charge
separation becomes $w_{RC}(t;\Psi_0)=2\kappa |{b}_{M+1}(t)|^2$, then we obtain an expression for the corresponding efficiency:
\begin{eqnarray}
\eta=2\kappa |B_0|^2 \int_{0}^{\infty}{\mathcal F} (t)dt \label{eq:eta}
\end{eqnarray}
with $B_0=\sum_{j=1}^M b_j(0)$. Equation (\ref{eq:eta}) is the main result of this paper. It shows that the efficiency
of transfer depends on the quantum coherence properties of the initial state as it becomes proportional to $|B_0|^2$ i.e. the {\it amplitude of
probability}, and not just the probability, that the excitation is initially in the LH1.
Therefore, the efficiency profile is sensitive to symmetric and asymmetric superpositions of localized excitation sates $|j\rangle$ i.e. it depends on the initial relative phases between states $|j\rangle$. From equation (\ref{eq:eta}) one can conclude that
symmetric delocalized excitation states yield an increase in $\eta$,
while some asymmetric states could be used to limit or even
prevent the transfer, i.e. $\eta=0$. Unless otherwise stated, we
henceforth consider symmetric initial states of the form
$|\Psi_{m}^{s}\rangle=(1/\sqrt m)\sum_{j=1}^m |j\rangle$ where $m\leq M$ is the
number of donors among which the excitation is initially delocalized. We denote $m$ the {\it delocalization length}.
For these symmetric states $|B_0|^2=m$ and hence $\eta \propto m$ as shown in Fig. \ref{fig:toy}(a). This figure also
shows that the efficiency  gradient
depends on strength of the interaction between one donor and the rest, which is quantified by $\Delta$. We have chosen
$\gamma$ to be the same for all these situations, but $J$ has been
taken to be such that $\Delta_{nearest}<\Delta_{dipole}\simeq
\Delta_{pairwise}$. For a fixed $m$, $\eta$ reaches higher values in the case of
nearest-neighbor couplings, while it achieves similar values
for dipole-dipole and pairwise interactions. According to these results, interaction among donors limits the
efficiency: the stronger the effective interaction between one donor and the ring, the lower the
efficiency will be. This phenomenon seems to resemble the `entanglement sharing' dynamics in the context of a central
spin coupled to a spin-bath \cite{dawson05}. In our case the LH1 complex can be seen as a spin bath for the RC. Dawson {\it et al.} \cite{dawson05} have discussed that interaction between bath spins translates to entanglement among them, and since {\it entanglement cannot be shared arbitrarily among several particles}, interaction among spins in the bath
limit entanglement between the central spin and the bath, and therefore may limit the efficiency of transfer.  A
discussion of efficiency in terms of entanglement is beyond the scope of this paper and the work in this direction
will be presented elsewhere.
Interestingly, $t_f$ in this simple model turns out to be independent of
$m$, as can be  deduced from Eq. (7). Therefore, for the symmetric
initial states considered, $t_f$ depends mainly on the mechanism of interaction, as can be seen in Fig.
\ref{fig:toy}(b). The decay-rate of $P(t;|\Psi_{m}^{s}\rangle)$ increases with $m$, as shown in Fig. \ref{fig:toy}(c).
Correspondingly, the excitation lifetime $\tau$ decreases as shown
in Fig. \ref{fig:toy}(d). The three situations satisfy $t_f\leq\tau$,
where the equality holds for the initial state in which the excitation is
symmetrically delocalized among all donors.
%%%%%%%%%%%%%%%%%%%%%%%%%%%%%%%%%%%%%%%%%%%%%%%%%%%%%%%%%%%%%%%%%%%%%%%%%
\subsection {A detailed model for the LH1-RC complex.}
%%%%%%%%%%%%%%%%%%%%%%%%%%%%%%%%%%%%%%%%%%%%%%%%%%%%%%%%%%%%%%%%%%%%%%%%%
We now apply the above formalism to the effective
Hamiltonian for the LH1-RC interaction given in references $9$ and $10$ \cite{hu97,hu98}.
In this case the RC has a special pair of BChl responsible for the charge separation,
i.e. $N=2$ acceptors, and two more accessory BChl molecules which do not participate in the charge separation process
\cite{ritz02} (see Fig. \ref{fig:psu}). The effective Hamiltonian is of the form given in equation (\ref{eq:hi}) but with certain particularities. First, the pigments at the RC are off-resonance with the donors. Second,
the interactions between adjacent molecules are quantified by two different
constants i.e. $J_{2j,2j+1}=\nu_1$ and
$J_{2j,2j-1}=\nu_2$ which are derived through quantum chemical calculations \cite{hu97,hu98}. Third, the coupling between non-neighbouring donors corresponds
to a dipole-dipole interaction of the form
$J_{jk}=\frac{{\vec \mu}_j\cdot {\vec \mu}_k}{r_{jk}^3} -
\frac{3({\vec  r}_{jk}\cdot {\vec \mu}_j)({\vec
 r}_{jk}\cdot{\vec \mu}_k)}{r_{jk}^5}$ where ${\vec \mu}_j$ is the transition dipole moment of the
$j^{th}$ donor and ${\vec r}_{jk}$ is the relative
position vector between donors $j$ and $k$. The directions of
${\boldmath \mu}_j$ have been taken from Hu et al.\cite{hu98}, and a top
view of the dipole representation of the LH1-RC core is shown in figure
\ref{fig:psu}(b).
%%%%%%%%%%%%%%%%%%%%%%%%  Fig 3 %%%%%%%%%%%%%%%%%%%%%%%%%%%%%%%%%%
\begin{figure}[tpb]
\centering
\resizebox{8cm}{!}{\includegraphics*{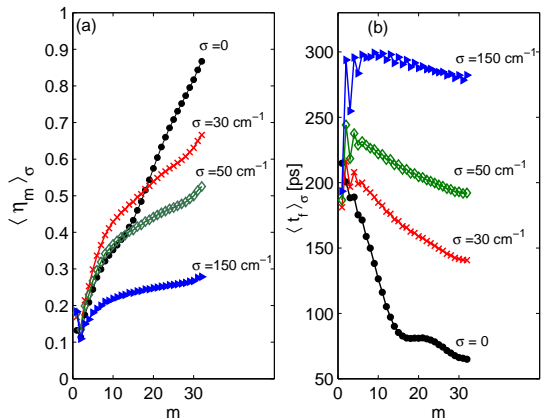}}
\caption{Numerical results for initial symmetric states of the
form $|\Psi_{m}^{s}\rangle=(1/\sqrt m)\sum_{j=1}^m |j\rangle$ satisfying $|B_0|^2=m$.
(a) Efficiency and (b) transfer time versus $m$ for different $\sigma-$values.
Results shown are averaged over an ensemble of $1000$ aggregates for each standard deviation $\sigma$.
Single-site energies and electronic couplings for the LH1-RC core have been taken reference \cite{hu98}.
For donors in the LH1 complex $E_{j}(0)= 12911 \ \rm {cm}^{-1}$
while for the special pair at the RC $E_{c}(0)= 12748 \ \rm {cm}^{-1}$, and for the accessory BChl.
$E_{a}(0)= 12338 \ \rm {cm}^{-1}$.
In all cases $\Gamma=1\ \rm{ns}^{-1}$ and $\kappa=4\ \rm{ps}^{-1}$.}
\label{fig:symm}
\end{figure}
%%%%%%%%%%%%%%%%%%%%%%  Fig 3 %%%%%%%%%%%%%%%%%%%%%%%%%%%%%%%%%%
\begin{figure}[tpb]
\centering
\resizebox{8cm}{!}{\includegraphics*{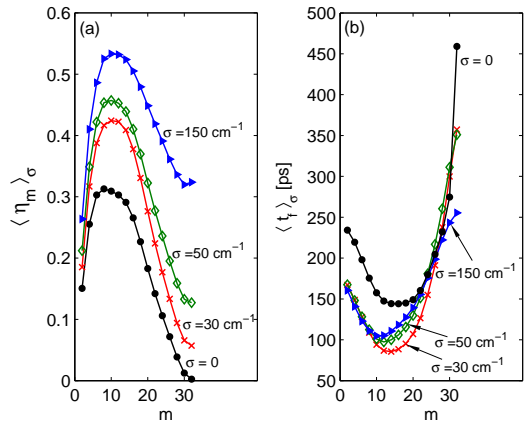}}
\caption{Numerical results for initial asymmetric states of the form $|\Psi_{m}^{as}\rangle=(1/\sqrt m)\sum_{j=1}^m
(-1)^j|j\rangle$ satisfying $|B_0|^2=0$. (a) Efficiency and (b) transfer time versus $m$
for different $\sigma-$values. Results shown are averaged over an ensemble of $1000$ aggregates for
each standard deviation $\sigma$. Single-site energies and electronic couplings for the LH1-RC core
have been taken from reference \cite{hu98}. For donors in the LH1 complex $E_{j}(0)= 12911 \ \rm {cm}^{-1}$
while for the special pair at the RC $E_{c}(0)= 12748 \ \rm {cm}^{-1}$, and for the accessory BChl.
$E_{a}(0)= 12338 \ \rm {cm}^{-1}$.
In all cases $\Gamma=1 \ \rm{ns}^{-1}$ and $\kappa=4 \ \rm{ps}^{-1}$.}
\label{fig:asym}
\end{figure}
%%%%%%%%%%%%%%%%%%%%%%  Fig 3 %%%%%%%%%%%%%%%%%%%%%%%%%%%%%%%%%%
For the initial condition of the excitation in the LH1 complex we consider both symmetric initial states
$|\Psi_{m}^{s}\rangle=(1/\sqrt m)\sum_{j=1}^m |j\rangle$ with total amplitude of probability $|B_0|^2=m$, and
asymmetric states $|\Psi_{m}^{as}\rangle=(1/\sqrt m)\sum_{j=1}^m
(-1)^j|j\rangle$ satisfying $|B_0|^2=0$. Since the system's dynamics is not invariant with respect to
pigment exchange, for each delocalization length $m$, we calculate the average efficiency
$\eta_m\equiv\langle\eta\rangle$ where the average is taken over all possible states for which the excitation is
delocalized among $m$ consecutive donors. Also, when energetic disorder is considered, for each value of standard
deviation $\sigma$, the efficiency corresponds to the ensemble average over $1000$ realizations of disordered $H$. We
denote this average efficiency $\langle\eta_m\rangle_{\sigma}$ and the corresponding average transfer time $\langle
t_f\rangle_{\sigma}$. An estimate of $\sigma=30\rm{cm}^{-1}$ for the standard deviation of the diagonal disorder
distribution in the LH1 has been provided in the literature\cite{timpmann05}. Therefore we have carried out calculations for $\sigma-$values ranging from $0$ to $150\rm{cm}^{-1}$  and the results for symmetric and asymmetric initial states are
shown in figures \ref{fig:symm} and \ref{fig:asym} respectively. For symmetric states and in the absence of energetic
disorder i.e. $\sigma=0$, the behaviour of the efficiency and transfer time are very similar to that in the toy model:
$\langle\eta_m\rangle_{\sigma=0}$ increases linearly with $m$ (see Fig. \ref{fig:symm}(a)) while $\langle
t_f\rangle_{\sigma=0}$ decreases (see Fig. \ref{fig:symm}(b)), achieving maximum efficiency and minimum transfer time
for the fully delocalized situation. For disorder distributions corresponding to small values of $\sigma$ i.e $\sigma
\leq 50\rm{cm}^{-1}$, there are symmetric states which exhibit efficiency improved in comparison to the
situation where no disorder is considered. Such states correspond to those for which $m<10$ as it can be seen in
figure \ref{fig:symm}(a). However, the efficiency values are clearly decreased with increasing $\sigma$.
Conversely for the asymmetric states satisfying $|B_0|^2=0$, the efficiency is
a non-monotonic function of $m$, indicating that there is an {\em
optimal} delocalization length for which $\langle\eta_m\rangle_{\sigma}$ has a maximum
 and for which $\langle t_f \rangle_{\sigma}$ has a minimum as it is shown in figure \ref{fig:asym}. The optimal
delocalization length is around $m=10$ as it can been see in figure \ref{fig:asym}(a). Such behaviour is robust to the presence of energetic disorder, and indeed improved as the efficiency values are larger with increasing $\sigma$.
It is also worth noting that in both cases, symmetric and asymmetric states, the efficiencies
obtained are lower than those given by a classical calculation with rate equations derived for the same
single-site energies, electronic couplings, and dissipation and charge separation rates here considered\cite{ritz01}.

The above results suggest that efficiency can therefore be used as an indicator for coherent energy transfer.
In particular, two-dimensional spectroscopy techniques recently developed to study coherence dynamics in photosynthetic systems\cite{engel07,lee07} may be used to create and probe quantum superposition initial states. For instance,
an optically allowed state of LH1 is the completely delocalized asymmetric state\cite{hu97}.
For very low values of energetic disorder and at very low temperatures, the efficiency of transfer from such state
is nearly zero. At a slightly higher temperature, the excitation will become less delocalized and as
such the efficiency would increase (c.f Fig. \ref{fig:asym}). In other words, an experiment measuring the
efficiency of an LH1-RC core under various temperatures, but still within a low temperature regime, could serve to ascertain the extend of coherence in energy transfer. Interestingly, some experimental works have indicated
that in thermalized LH2 complexes the excitation may be coherently delocalized over just
a few donors of the B850 ring\cite{trikunas01} while at very low temperatures it can be fully delocalized over the whole ring \cite{oijen99}. Unfortunately, no such investigation has been reported on the LH1.

In conclusion, we have presented a formalism to study the role of quantum coherence in photosynthetic
units that exhibit coherent energy transfer mechanism. Our results open up experimental possibilities to investigate
and exploit such coherent phenomena in artificial and natural systems harvesting light.

\acknowledgments
A.O-C thanks Trinity College (Oxford) for financial support. C-F.L thanks Glasstone Trust (Oxford) and
Jesus College (Oxford) for financial support. F.F.O is grateful to Conicyt (Chile).  

\appendix
\section{Details of the analytical solutions for the Toy Model}
As described in section III (A), in the simplest situation where the RC is taken as a single two-level system
on resonance with $M$ donors, we are able to find analytical solutions for the probability density of having a
charge separation event at the RC, and therefore we find an analytical expression for the efficiency of transfer
in this toy model (see Eq.(\ref{eq:eta})). Here we give the details of this calculation.
The unnormalized state $|{\Psi}_{cond}(t)\rangle$ given in equation (\ref{eq:state}) satisfies the relation
$d|{\Psi}_{cond}(t)\rangle/dt=-iH_{cond}|\Psi_{cond}(t)\rangle$ which leads to the following set of first-order
coupled differential equations
{\small
\begin{eqnarray}\label{eq:bj}
\dot{b}_j(t) & = &-i \gamma_{j}b_{M+1}(t) -i \sum_{k\neq j}^{M} J_{jk}b_k (t) -\Gamma b_j (t)\  \\ \label{eq:bc}
\dot{b}_{M+1}(t)& = &-i \sum_{j=1}^M \gamma_{j}b_{j}(t) -\kappa b_{M+1}(t)\ .
\end{eqnarray}}
Assuming that all donor-RC coupling are identical i.e. $\gamma_{j}~\equiv~\gamma$ for all $j=1\dots 32$, we have
\[
\ddot{b}_{M+1}(t)+\kappa \dot{b}_{M+1}(t)+M\gamma^2 b_{M+1}(t)=\mathcal{G}(t)
\]with
\[
\mathcal{G}(t)=i \gamma \Gamma \sum_{j=1}^{M} b_j (t) - \gamma\sum_{j=1}^{M}\sum_{k\neq j}^{M} J_{jk} b_k (t) \ .
\] Defining the effective coupling between one donor and the LH1 ring as $\Delta_j=\sum_{k\neq j}^{M} J_{jk}$, and assuming it identical for all the donors in the LH1 i.e. $\Delta_j\equiv\Delta$, we have
$\mathcal{G}(t)=\gamma (i\Gamma-\Delta) \sum_{j=1}^{M} b_j (t)$. Now from equation (\ref{eq:bc}) we get $\sum_{j=1}^{M} b_j(t)=i\gamma^{-1}(\dot{b}_{M+1}(t)+\kappa b_{M+1}(t))$ and thus we arrive to
\begin{eqnarray}
\ddot{b}_{M+1}(t) + X \dot{b}_{M+1}(t) + Y {b}_{M+1}(t) = 0 \ ,
\end{eqnarray}
with $X=(\kappa+\Gamma +i\Delta)$ and $Y=4M\gamma^2 +\kappa (\Gamma+i\Delta)$.
The solutions of the above differential equation are of the form
\begin{equation}
b_{M+1}(t) = f(t)\sum_{j=1}^{M}b_j(0)+ g(t)b_{M+1}(0)
\end{equation}with \begin{eqnarray*}
f(t) &=&-2i\gamma e^{-Xt/2} \frac{\rm{sin}(\Omega t/2)}{\Omega}\\
g(t) &=&e^{-Xt/2}\left[\frac{(\Gamma-\kappa +i\Delta)}{\Omega}\rm{sin}(\Omega t/2)
+\rm{cos}(\Omega t/2)\right]
\end{eqnarray*}
and $\Omega=\sqrt{4M\gamma^2-(\Gamma-\kappa+i\Delta)^2}$. The above solution is invariant
with respect to exchange of any pair of two-level systems. It is also worth noting that the above formalism allows
to find a closed expression for the collective amplitude of probability $B(t)=\sum_{j}^{M}b_j(t)$ which leads to further
simplifications of this system.\cite{olaya}

\end{document}